\documentclass[sigconf, nonacm]{acmart}
\usepackage{rotating}
\usepackage{xcolor}
\usepackage{color}
\usepackage{colortbl}

\definecolor{Gray}{gray}{0.98}

\AtBeginDocument{%
  }

\begin{document}
\thispagestyle{empty}

\title{Stepping into the Margins: How Readers Want AI to Generate Footnotes}


\author{Piper Vasicek}
\affiliation{%
  \institution{Brigham Young University}
  \city{Provo}
  \state{Utah}
  \country{USA}}
\email{piper.vasicek@gmail.com}

\author{Courtni Byun}
\email{cbyun22@byu.edu}
\affiliation{%
  \institution{Brigham Young University}
  \city{Provo}
  \state{Utah}
  \country{USA}
}

\author{Kevin Seppi}
\affiliation{%
  \institution{Brigham Young University}
  \city{Provo}
  \state{Utah}
  \country{USA}
}


\begin{abstract}
Footnotes can be powerful tools to aid understanding, providing information that augments the reading experience. However, static footnotes cannot address every reader question. Current reading tools allow readers to view curated footnotes, allow personal and social annotation, and link dictionaries to reading material. Many other existing tools and natural language processing (NLP) techniques--such as generative AI, summarization and translation--could be used to address any reader question. However, no one has yet explored which of these features readers actually want. To bridge this gap, we conducted thirteen semi-structured interviews with readers from various backgrounds, followed by a thematic analysis of their responses. We develop themes describing the types of footnotes readers prefer and how to determine the quality of footnotes--specifically focusing on what sources of information a system considers, what the footnotes contain, and how the footnotes are presented to the reader.
\end{abstract}





\maketitle

\section{Introduction}

Self-directed learners learn best when they deeply engage with the text they are reading. One method of engaging with the text is through footnotes. Footnotes--informational asides included at the bottom of a page--have existed for hundreds of years, and may include further context, citations, word definitions, cross-references, commentary, translation, and even images~\cite{history}.

While footnotes as they currently exist are useful, they have certain limitations. For many nonfiction books, footnotes are available as included by the author. Otherwise, curated footnotes are available only for widely studied texts, such as religious texts and well-known works of classic literature. Even when curated footnotes are available--whether from the author or another source, it is not possible for those footnotes to address every possible point of interest--a footnote may not exist for a passage of interest, or else the available footnotes for that passage may not apply to the reader’s particular interest about that passage.
There is simply no way for a static set of footnotes to contain all related information that a reader might desire.

Even if such a resource could be created, it would prove unwieldy. While often useful, footnotes can be distracting to readers, particularly when those footnotes do not lead to information the reader is interested in. Indeed, a long debate exists in the law community about whether footnotes should exist at all, with some arguing that they are an unwelcome diversion, and that all relevant information should be included in the body of the text~\cite{bye_footnotes,bye_law_reviews,defense_footnotes}. In reference to footnotes being included in play scripts, John Barrymore famously quipped, ``It's like having to run downstairs to answer the doorbell during the first night of the honeymoon''~\cite{barrymore}.
 
To some extent, collaborative annotation can alleviate the issue of completeness and interest.
Annotation--adding comments to a text--has existed nearly as long as the written word, with ancient manuscripts 
including hand-written notes added in the margins~\cite{scholia}. 
The digital age has produced tools for collaborative annotation where a reader may flag any passage or topic with questions, and a collaborator may share insight into those questions. These annotations are relevant to the initial annotator, and hopefully useful to the collaborator as well. However, the usefulness of a response to any particular question is highly dependent on the group collaboratively annotating, and collaborators may take a significant amount of time to respond to a particular question from the reader or not reply at all.

A potential solution to this problem would be to find a way to implement some form of AI annotation. Many natural language processing (NLP) tools now exist that can be used to elucidate the content and meaning of texts, e.g. automatic text summarization, topic modeling, automatic cross-referencing, large language models (LLMs), text-to-image models, etc. 

These AI tools could respond almost instantaneously to readers and could be asked to provide commentary on any portion of the text desired. They could be implemented in conjunction with curated footnotes and collaborative annotation tools—if desired and available--but could also be applied to obscure texts that lack the support of other readers and annotators.
Indeed, the AI features that could be integrated into an e-reader or other digital reading environment are practically limitless. And that is precisely the challenge.

Before such tools are created and implemented, it is important to understand the current behavior and desires of readers in relation to footnotes and annotations.

We consider the following research questions:

\begin{itemize}
    \item RQ1: What footnote tools, if any, do readers currently use (specifically for reading)?
    \item RQ2: When selecting text, what would readers want to see from an AI footnote; does it depend on the type of text being read?
    \item RQ3: What would ideal AI footnotes look like?
\end{itemize}

In order to come to a better understanding of readers' desires and expectations, we conducted semi-structured interviews with 13 readers from a wide variety of reading backgrounds regarding footnotes and their thoughts about AI footnotes on-demand. We then conducted an inductive thematic analysis of their responses and developed themes around the types of footnotes readers want and how to judge footnote quality, offering a conceptual foundation for designing functioning AI footnote systems.

\section{Related Work}

We define a few terms here for the sake of clarity.

\textbf{Footnotes} are asides added to a text by the author or editor of the text. In this paper we do not distinguish between footnotes, endnotes, and intra-textual commentary. Our goal is to understand how to best support reader understanding. Therefore, we use the term footnotes to include any notes, commentary, citations, etc. that are not part of the main body of the text, but that are meant to elucidate the main text in some way.

\textbf{Social annotation} involves a group collaboratively commenting on a text. There has been a significant amount of work on the benefits of social annotation. Many of the studies involving social annotation are case studies~\cite{sa_case_study_3,sa_casestudy_1, sa_casestudy_2}, explore the benefits of collaboration~\cite{sa_benefits_1,sa_benefits_2,sa_benefits_3}, or evaluate the usefulness of social annotation for particular tasks~\cite{sa_task_1,sa_task_2,sa_task_3}.


The ability to answer reader questions in real-time has led to an interest in developing an ``intelligent textbook'' which a reader can ask questions~\cite{inquire-biology,textbook-survey,interactive-features}. A 2020 paper explores an intelligent textbook which uses a knowledge graph and text comparison to answer these reader questions~\cite{AI-biology}. However, rather than answering a reader's question directly, the system matches the question to questions prepared using the expert-created knowledge graph, and the readers are therefore still limited to a predetermined set of subjects about which they can inquire.

Other augmented reading tools have also been produced. 
Semantic Scholar provides a reading environment with citations that pop-up in text 
and AI-driven mark-up for skimming papers~\cite{semantic-reader}. Many e-readers incorporate ways to view word definitions as you read~\cite{ereader-dictionary}. Enhanced eBooks exist that incorporate animations, and even haptic or olfactory effects--if almost exclusively in prototype form~\cite{multisensorial, animations-children}.

Significant research has gone into automatic citation for the research community~\cite{citeseer,cite-recommend-survey}. Recent work in source attribution has employed LLMs to rank possible sources for claims~\cite{source-attribution} or has used LLMs for literary evidence retrieval--the ability to find a relevant quote based on given commentary~\cite{literary-claims}.
There has also been some investigation into using topic modeling to produce cross-references--a specific type of footnote~\cite{xref-og}.

On the other hand, there has been no work, to our knowledge, on producing discursive footnotes--as opposed to citational footnotes--despite the fact that many NLP tools exist that could be applied to footnotes in that way. It is also unclear to what extent available footnote-related tools align with what readers actually want. To bridge this gap, we ask the readers themselves.

\begin{table*}[tbh!]
    \vspace{2cm}
    
\renewcommand{\arraystretch}{1.3}
    \begin{tabular}{lccccccccclcccccccccclcccc}
        &
        \begin{rotate}{75}Genealogy \end{rotate} & 
        \begin{rotate}{75}Geology \end{rotate} &
        \begin{rotate}{75}History \end{rotate} & 
        \begin{rotate}{75}Law \end{rotate}&
        \begin{rotate}{75}Linguistics\end{rotate} & 
        \begin{rotate}{75}Primary Sources \end{rotate} & 
        \begin{rotate}{75}Religion\end{rotate} & 
        \begin{rotate}{75}Technology \end{rotate}&
        \begin{rotate}{75}Unspecified\end{rotate}&
        \phantom{i} &
        \begin{rotate}{75}Art\end{rotate}&
        \begin{rotate}{75}Biography\end{rotate}&
        \begin{rotate}{75}Creative Nonfiction \end{rotate}& 
        \begin{rotate}{75}Do-It-Yourself \end{rotate}& 
        \begin{rotate}{75}Finance\end{rotate}&
        \begin{rotate}{75}News\end{rotate} &
        \begin{rotate}{75}Psychology \end{rotate}& 
        \begin{rotate}{75}Religion \end{rotate}& 
        \begin{rotate}{75}Self Help \end{rotate}& 
        \begin{rotate}{75}Social Science \end{rotate} & 
        \phantom{i} &
        \begin{rotate}{75}Classics \end{rotate}&         
        \begin{rotate}{75}Foreign Language \end{rotate}& 
        \begin{rotate}{75}Literary Fiction \end{rotate}&
        \begin{rotate}{75}SciFi/Fantasy \end{rotate} \\
        \cmidrule{2-10} \cmidrule{12-21} \cmidrule{23-26}

        & \multicolumn{9}{c}{Academic Papers} &
        & \multicolumn{10}{c}{Nonfiction} &
        & \multicolumn{4}{c}{Fiction}\\
        \toprule

\rowcolor{Gray}P1&&&&X&&&&&&&&&&&&&&&&&&&&&\\
P2&&&&X&&&&&&&&&&&&&X&&X&&&&&&\\
\rowcolor{Gray}P3&&&&X&&&&&&&X&X&&&&&&X&&&&&&&\\
P4&&&&&&&&&&&&X&X&&&&&X&&&&&&&X\\
\rowcolor{Gray}P5&&&X&&&&X&&X&&&&&&&&&X&&&&&&X&\\
P6&&&&&&&&X&&&&&&&&&&X&&X&&&&&X\\
\rowcolor{Gray}P7&&&&&&&&&&&&&&&X&&&X&X&&&&X&&\\
P8&&&&&&&&&X&&&&&X&&&&X&&&&X&&&X\\
\rowcolor{Gray}P9&&&&&&&X&&&&&&&&&&&X&&&&&&&\\
P10&&X&&&&&&&&&&&&&&&&X&&&&&&&\\
\rowcolor{Gray}P11&&&&&X&&&&&&&&&&&X&&X&&&&&&&X\\
P12&X&&X&&&X&&&&&&&&&&&&X&&&&X&X&&X\\
\rowcolor{Gray}P13&&&&&&&&X&&&&&&&&X&&X&&&&X&&& \\

    \end{tabular}
    \caption{Interview participants listed with their preferred genres, including academic fields of study and hobby reading. Participants read or studied a median of 4 different genres.}
    \label{tab:genres}
\end{table*}

\section{Method}
We conducted interviews about current footnote use as described below. 
All procedures were approved by our institutional review board.

\subsection{Participants}

We conducted semi-structured interviews with 13 footnote users aged 23 to 79 ($\mu=35$, 7 male, 6 female). Participants were recruited through the authors' contacts and snowball sampling. We attempted to find participants from a variety of fields with varied experience using footnotes.
Table~\ref{tab:genres} shows the genres which each participant frequently reads or studies.

Our participants tended to be highly educated. All of them have completed at least some college, and 4 of them have doctoral degrees.
The specific population we interviewed also tended to be more religious--most of our participants use footnotes for personal religious study in addition to using footnotes in their career/main field of study, and many are also hobby readers.
While these traits make our participants a less representative sample of the general population, we consider it advantageous to the current study that they have extensive experience with footnotes in multiple capacities. We asked them about footnote use and what they wanted from an AI footnote system in each of their various reading interactions. 

\subsection{Interviews}
We performed interviews in person or via video conference, based on participant preference. The lead author conducted all interviews. Interviews were semi-structured with questions about reading habits, current footnote use, and how AI could be integrated into footnotes. Participants received a \$15 gift card for participation.

In-person interviews were audio recorded, and video conference interviews were audio and video recorded. The audio was transcribed by Otter.ai\footnote{\url{https://otter.ai/}} and the transcripts were reviewed, corrected, and deidentified by the lead author who conducted the interviews.

\subsection{Data analysis}

We conducted an inductive thematic analysis as described in~\citet{rta-1}.

Two authors read through all of the interview transcripts. After reading through them, the authors developed 61 initial codes. Over time we combined these into four overarching codes (Text Types/Contexts, Footnote Types, Indicators of Quality, and Attitudes about AI). 
We then individually applied these codes to the transcripts on a sentence-by-sentence basis. Two of our authors met several times to reconcile the codes, holding out the third author to act as an arbitrator, if necessary. It ultimately proved unnecessary as the two authors were able to reconcile the codes without a tiebreaker. Because this is an inductive thematic analysis, we do not report inter-rater reliability~\cite{to-irr-or-not}.

We further considered all text coded with \textbf{Footnote Types} and \textbf{Indicators of Quality} codes. We developed five overarching themes for types of footnotes. We further developed three themes linked to determining the quality of AI footnotes. 

\section{Findings}
We considered interview responses from the perspective of developing a conceptual foundation for implementing and evaluating AI footnotes. To this end, we developed themes surrounding types of footnotes readers want (which relates to RQ1 and RQ2) and about how to evaluate footnote quality (which relates to RQ3).

\subsection{Types of Footnotes}
\begin{quote}
    \textit{RQ1: What footnote tools, if any, do readers currently use (specifically for reading)?}
    
    \textit{RQ2: When selecting text, what would readers want to see from an AI footnote; does it depend on the type of text being read?}

\end{quote}

\begin{figure*}
    \centering
    \includegraphics[width=1\linewidth]{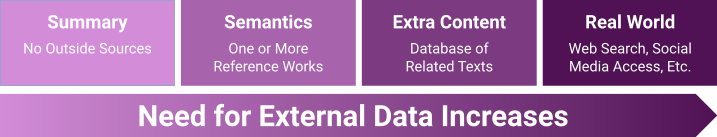}
    \caption{Four broad types of footnotes ordered by how much external data would be required to produce them. 
    }
    \label{fig:data_needs}
\end{figure*}

In answering our first two research questions, we developed participant-suggested types of footnotes into five overarching themes based on the amount of information external to the text that would be needed to produce them. We represent this graphically in Figure~\ref{fig:data_needs}.

\subsubsection{Summary of given text (No Outside Sources)}
The summary theme includes types of footnotes that can be produced using only the text that the participant is currently reading. Four different types of footnotes were suggested in this category: story recaps, character bios, consistency checks, and rephrasing of complex ideas. The majority of the first three were discussed by participants in the context of reading or writing fiction, while the final category was mentioned in the context of academic reading and writing. 

\paragraph{1. Story Recap}
P11 suggested the ability to get a recap of what had already been written in a fiction setting: ``Sometimes when I pick up a book, I read it for a little bit and I come back to it later. And so if I'm reading the text that way it could be beneficial to be, like, `Here's a recap of maybe some of the, like, ties between, like, oh, this character has appeared through this, you know, this kind of storyline. And these are some of the things you may have forgotten.'\,''

\paragraph{2. Character Bio}
Similarly, P12 suggested a system with the ability to produce character bios: ``So it's like, `Oh, Shalan, who's Shalan?' Boom, there's a character bio.''

\paragraph{3. Rephrase}
In law and other academic papers, participants believed AI could be useful to rephrase difficult concepts to make them easier to understand. P2: ``I feel like maybe just like rewording or explaining things that were confusing.''

\paragraph{4. Consistency Check}
We are not concerned with AI tools for writing in this research; however, it is interesting to note that participants often conflated using footnote tools for reading and using footnote tools for writing. We therefore include that for fiction, P6 suggested that AI might be capable of checking for continuity errors as a support for fiction writers: ``...maybe as a consistency checker would be useful, but also to remind you--remind me `Oh, this character's eyes are blue. You just put green here.'' 


\subsubsection{Semantics (One or More Outside Reference Sources)}
Many participants suggested that AI could be useful to connect to authoritative reference sources, like various forms of dictionaries (e.g. usage dictionaries, etymology dictionaries, etc.).

\paragraph{1. Context-Aware Definitions}
Most participants (P4, P5, P6, P8, P9, P11, P12, P13) wanted an AI to have the ability to provide definitions. 
These participants suggested that AI might be able to give more contextually valid definitions of terms, which could apply to religious texts, academic papers, and fiction. Several varieties of reference works were mentioned as potentially being useful, including \textit{foreign language dictionaries}, \textit{etymology dictionaries}, and \textit{field-specific technical dictionaries}. P8: ``If it's a word, I would at least want a definition of it. And especially in religious study, I want the etymology of the word and history of it. And possibly, if it was translated from something else, and that etymology and history.'' P10 emphasized the need for field-specificity: ``If you get a geophysical term, it's almost always going to--when you go search on it, it's almost always going to take you off on to the physics end of chemistry or something. And the same with petrology. And it's not helpful because it's not specific enough to what you're thinking about. More general physics, more general chemistry, what you need to know is what that means in the geological context.''

\paragraph{2. Concept Prioritization}
In a scriptural context, P13 suggested that AI could determine, for a given concept, ``how central is this to the set of however many core doctrines there are in the associated religion?'' In other words, AI could be used to indicate to the reader how important it is to understand a particular concept in a given field.


\subsubsection{Extra Content}
\paragraph{1. Supplemented Citations}
Every participant who reported reading in an academic field mentioned the need for citations. Their preference was nearly always to include the relevant portion of the text in addition to the citation, so that they could evaluate the source without having to track it down. P11: ``Like, actually, this book, \textit{Distant Reading}, right here is referenced by a ton of people. But it's more like the overall concept, and there's not really any passage that like is actually helpful to what I'm doing. So I'm like, `Okay, well, I didn't need to read the book.' So maybe even just like some summary, like adding to annotations that are already there, like being able to add like, `Oh, this is the abstract for this article, so you don't have to go and find it yourself.'\,''

\paragraph{2. Similar Recommendations}
Participants also requested the ability to find similar passages of text from other sources in academic reading--P5: ``So if there was some way that AI could...kind of get a general sense for the kinds of words and the kinds of constructions that are common in this text, and then like, suggest other works of fiction that have similar, kind of, global vocabulary or maybe global style or something like that, that could be really fun in a comparative way, right?''

A similar idea was requested for hobby reading--P8: ``If I can look up a scene in a book--like a paragraph in a book that I'm enjoying reading, going, like: `I wonder what other novels--' and this would be large groups of data, but ``--what other novels have a scene like this?' That would be awesome, because that would be new reading material right there.''

\paragraph{3. Commentaries}
Several participants speaking about religious texts, whether from an academic point of view or for personal study, requested the ability to link to commentaries on scriptural passages (P3, P5, P8, P10). 



\subsubsection{Connection to the ``real'' world}
In a wide variety of genres, many participants wanted historical or cultural content from the real world to give context to a given passage.

\paragraph{1. Real-World Events}
Many participants suggested that AI footnotes could connect what they were reading to the world outside their reading. For law, P2 suggested that AI could inform the reader if current events in the supreme court or Congress might affect a particular case ``...if the AI could tell you that a case was like, coming before the Supreme Court or had just applied to the Supreme Court, and like, could be decided later. And we were like, waiting on or maybe like, we're waiting on a decision, that would be super cool. And people would be obsessed with that.''

For fiction texts, multiple participants suggested the ability to tell them when a book--contemporary or classic fiction--was referencing a real-world event. Examples vary from references in fantasy humor novels by Terry Pratchett (P8) to Victor Hugo (P12).

\paragraph{2. Real-World Discussions}
Finally, for both company financial disclosures, and scifi/fantasy, participants suggested that AI could link them to online discussion forums. In the case of company financial disclosures P7 requested, ``something that I could check to see what everybody is saying about this particular disclosure. I mean, that would help me to not have to do as much of my own research as I might do. Because if everybody's saying the same thing, you know, using AI, that's going to speed up my time of doing research.''
In fiction, P4 and P12 suggested linking to fan-made wikis, Reddit threads, and Discord servers for extra insight into fiction books with large followings. This particular kind of footnote seems to fill the role that social annotation--groups collaboratively reading a text--fills, 
where the reader seeks to understand a text based on the thoughts of interested parties.


\subsubsection{Multimedia}
Finally, only four of the participants mentioned going beyond text into other multimedia forms in the footnotes. This may be because current footnotes are almost exclusively text-based, and so thinking about image- or video-based footnotes requires a creative leap. Despite the fact that multimedia footnotes were not mentioned frequently, we discuss them here because much of the focus for augmented reading has historically been on animations, illustrations, etc.

\paragraph{1. Data Visualizations}
P10 talked about figures in academic papers as a form of annotation: ``You could kind of think of the figures as a type of annotation, because of how they're used.'' And P8 suggests that it would be useful if the AI could aggregate and re-display data in books with a lot of tabular data: ``And especially if you can compare portions of that table with other tables with the same kinds of data, instead of having to switch back and forth with a find function. That would be really, really useful.''

\paragraph{2. Video}
P4 suggested the ability to link to video content in order to explain complex academic topics: ``if I'm learning about critical race theory, then, like, here's a three-minute video that, like, summarizes critical race theory.''

\paragraph{3. Images}
P8 suggested being able to produce pictures of real-world objects: ``[My daughter's] going to be reading the classics starting in not that long. I would love to have a tool like this, where I can say, ``...Here's what a steamboat looks like.'\,''

\paragraph{4. Fan Art}
P12 suggested the ability to link to fan art for scifi/fantasy in Reddit/Discord fan groups of fan wikis: ``Here's, like, six different artist interpretations of what she looks like, you know, that are actually based on Brandon's textual description of it, right?''


\subsection{Questions of Quality}
\begin{quote}
    \textit{RQ3: What would ideal AI footnotes look like?}
\end{quote}
As someone implementing AI footnotes how can we determine if the footnotes are quality footnotes? Participants discussed what makes a footnote useful in various contexts. In considering our final research question, we develop themes that relate questions of quality to the sources used to train or be accessed by the AI, formulating the footnotes, and presenting footnotes to the reader. 

\subsubsection{Sources}

The central ideas related to source quality were that AI-recommended sources need to be credible in the context in which they are being used and that care needs to be taken to avoid AI-recommended sources leading the reader into an echo chamber.



\paragraph{1. Credibility and Pitfalls}
Participants mentioned the potential biases found in sources, and the credibility of various sources. They say that the credibility of a source depends on the context in which a source is being used. P4 expressed the idea that what is credible differs based on particular genres and from professional capacity to hobby reading: ``In my research setting, maybe I don't have to explain that like I can't use something on like a personal blog. I can't use something on Wikipedia that doesn't have a source. Like, it knows what a reliable source is. It knows maybe what type of like, maybe it knows that I have access to my institution to like the JSTOR database or whatever, like the different resources I have, and it can pull from that type of thing where when I'm on like, my pleasure reading setting or whatever, it's like `Oh, like, we're less concerned about like, Wikipedia articles. So, I found like, this Reddit thread that, like, talks about this fan lore.'\,'' 

P10 pointed out that even within a single genre, authoritative sources may sometimes have blind spots, sources that are generally considered less credible can supply good information in certain contexts. In other words,  the credibility of a source may need to be considered on a case-by-case basis: ``When it's going out for outside sources can it distinguish--I mean, when you--when it comes to a lot of fields like biblical archaeology, eschatology, there's a lot of pitfalls. There's people who have kind of outer space alien kind of theories. So that can be a pitfall. Sometimes the stuff in there is actually worth looking at, because they've gone and done the homework to say, `Oh, well, this city was next to that city.' If you look at the, kind of the more classical writers or the more current academic writers, they have their biases, you know, so that they fail to see things that should be obvious. I think people who write about the New Testament, the prejudice now is that they consistently misread the Old Testament. And so, they underplayed links to the Old Testament. And so, and there's many other blind spots like that, that the scholarship is fraught with. And so--like, when you're looking at Google search results they'll often give you a list of, you know, questions you might ask, and you click on that, and it takes you to the answer, right. And the answer is usually a segment of a paragraph from a Wikipedia article or some similar source. And it is an answer, but there's lots of other answers that might be more authoritative or better. So can it tell that 'Oh, well, this is the kind of source I favor.' That's, that's a hard one. But it'd be great if it could do it because it would save hours and hours and hours.'' 

\paragraph{2. Avoiding Echo Chambers}
Participants also mentioned that weeding out opposing opinions may lead to being stuck in an echo chamber. P5 discussed this: ``I worry about...the Google bubble, right? The news bubble that people talk, the ideological bubble that people talk about, I don't want it to start serving me only what it thinks I want to see. Because I think that encountering things that you don't want to see, or that you didn't even know that you didn't know, right, is a really important part of scholarly discovery. So I guess I would be hesitant to want it to become again too convenient or too attuned to what it thinks that it--that I want, or even what I think that I want, because then I might miss something that could be really generative.''

Related to the idea of an echo chamber, P12 pointed out that AI may be able to find sources to support any opinion: ``And then, yeah, if you're trying, like, if I were to write a paper or something, and then I wanted to use AI to like, go find me sources, you know, that there could also be confirmation bias in a way. Like AI is like, `Okay, this person has written this which is basically amounts to an opinion. And you could probably find sources to reinforce almost any opinion. And here you go, there they are.' And now you've got what looks like a very, you know, like a thoroughly researched paper, but it's really still just an opinion essay, where they happened to find things that agreed with you and ignore the ones that don't.''



\subsubsection{Conception: the Goldilocks Principle}

In actually formulating footnotes, participants focus on the Goldilocks principle: does the footnote contain the information the reader needs without including extraneous information. 

\paragraph{1. Footnote Length}
One aspect of this is the length of a footnote. P10 addressed this: ``It'd be great if you could get information on historical background that was tailored. That's the key, because you can always find information. What you want is relevant information. So for example, I believe that if you are reading the D\&C, and you say, ``Well, here's a Jason Coalville or something.'' You--what you want to know of him is limited to about two sentences, maybe three, not six. Because you don't want to interrupt the flow of your reading, but you want to know enough about him to say, is there any context here?''

This can also be seen in a tension between supplying the reader with all possible options for related material vs a more curated list that can predetermine what will be most useful. P9 talks about the issue of information overload: 
``[W]here do you stop? You know, especially in certain fields. I mean, on church stuff if you bring up atonement, good luck, you know. And how do you sort through that? Just that mountain of information that's going to come and where do you stop?''

\paragraph{2. Filtration}
P13 also emphasizes the need for filtration: ``just the presence of the idea in the corpus wouldn't mean it's a good idea. We'd need to then have some sort of value judgment as to is this really authoritative? Has it been supported or is it one of these things that's been diminished?'' However, this is complicated by the fact that there are certain conditions under which the reader does simply want a list of all possible information. P13 speaks to this in a scriptural context: ``It'd be interesting for--in that kind of context for the AI to say, 'Well, here's the comprehensive list of other scripture citations that you could go look at in reading the scriptures.'' 



Participants also discussed how they might like to have control over the filtration process of an AI footnote system. P8 suggests: ``If I had a--and this would be an on-the-fly thing, it would be amazing if I could say I want layers of annotation, where it's, like, this is the bare basics, the minimum needed, and above that, and maybe like three levels, where the third is like every single thing I can control click or something, and get that word or phrase or whatever [within the current footnote] looked up, that would be amazing.''





\subsubsection{Presentation}

AI footnotes carry the same burden of user design that applies to all new technologies--ensuring that the benefits outweigh the drawbacks of learning to use it. AS P5 said ``I have a low tolerance for, like, learning new programs.''

\paragraph{1. Passivity vs Intrusion}

Participants also mentioned the potential for an AI footnote system to be intrusive and expressed a hard line against this. P6 said that they would not use AI annotation ``if it was over-intrusive and wanting to guide me in a certain way...Yeah, like, if you're reading the scriptures, you don't want this AI thing determining your scripture study, for example. I think a lot of people wouldn't like that. But if it's just like a passive thing on the side, so `Oh yeah, you can look at this if you want that or this other thing that gives a little more information,' then I--yeah, I don't see why anyone wouldn't want to use it.''


\section{Discussion}
Participants mentioned a variety of footnote types that would be important for an AI footnote system to be able to generate. They also mentioned different features that could indicate a good quality footnote. We will now discuss design implications that follow from the footnote types and quality measures participants mentioned.

\subsection{Designing AI Footnotes: Different Types and Data Sources}
Participants mentioned wanting footnotes that would require varying degrees of external information. The types of footnotes they mentioned were summaries, which would require no external sources, semantic footnotes, which would require one or more external sources, extra content, which would require a database of related texts, and real world footnotes, which would require web search results and social media, among other sources. 

\subsubsection{Summary Footnotes}
Participants mentioned four different types of summary footnotes they would like to see in an AI footnote system: story recaps, character bios, consistency checks, and rephrasing of complex ideas. Each of these features could theoretically be implemented while only referring to the current text and without reference to outside sources. Many currently available techniques can be applied to this problem including the current state-of-the-art technology in summarization~\cite{summarization}, although implementers may need to consider how to summarize without spoiling the ending. For example, it may be necessary to track which characters and events have already been introduced into a novel, so an AI footnote system can be sure to only discuss those that have already been introduced. However, it is worth noting that such a feature may be most desirable for a casual reader's first reading of a novel, while in an academic setting having information about how a certain topic fits in with later material may help a student better contextualize what they are studying. 

Many other niche text processing techniques could also be brought to bear that only look at the text, such as plot extraction for fiction or biography~\cite{plot_extraction}.




\subsubsection{In-Context Clarifications}
Another type of footnote participants were interested in was definitions provided in the context of the text. E-readers often have the ability to link to a dictionary in order to define words and most participants wanted an AI that could provide word definitions.
Each of these types of footnotes would require access to reference works at a minimum. However, being able to define words in context may take access to more specialized algorithms, such as LLMs, in order to produce a nuanced definition appropriate to the given field. It may even be necessary to give the LLM access to a specialized database in the field and implement some form of Retrieval Augmented Generation (RAG)~\cite{RAG} to retrieve satisfactory results at this stage.

\subsubsection{Supplemental Related Content}
A footnote system that could provide citations with relevant portions of text, similar passages from other academic sources or other works (of fiction) that have similar vocab or style, other novels with similar scenes, links to commentaries on scriptural passages, etc., would require access to potentially vast amounts of text to be able to perform the comparisons requested or to provide excerpts to the reader. In this case, RAG might also be useful~\cite{RAG}. It could also be useful to consider implementing work in literary evidence retrieval, cross-reference generation, or source attribution~\cite{literary-claims,source-attribution,xref-og}.

Regardless of the method employed to find useful passages, some form of database would be necessary to provide text passages. If it is undesirable to maintain such a database for the AI footnote system, it may be possible to use an available API with access to vast amounts of text, such as the Google Books API\footnote{\url{https://developers.google.com/books}} or the Semantic Scholar API\footnote{\url{https://www.semanticscholar.org/product/api}}. Other more niche databases may also be available for specific applications.

\subsubsection{Making ``Real'' World Connections}

Participants indicated a desire for footnotes that provide context on events in the world of the text, be that real or fictional. These could include background information for real or fictional historical events, or links to real-world discussions relevant to the text. In all cases, these types of footnotes would require the ability to find and synthesize content potentially from a wide variety of sources.
This might mitigate the need to maintain a large database of text specifically for the system; however, it might also require other means of accessing large amounts of data such as through web search and social media APIs. 
It may also require more niche algorithmic solutions that apply to niche areas for finding historic and cultural connections, especially for obscure works.

\subsubsection{Multimedia Footnotes}
Participants discussed a desire for footnotes that included images, videos, fan art, or data visualization tools. The algorithmic necessities for each of these kinds of multimedia footnote are broad and varied. Each would need to be considered on a case-by-case basis. Obviously, text-to-image generators could be employed to produce images on-demand~\cite{text-to-image}. Even on-demand video generation is becoming a possibility~\cite{video-generation}. We must, therefore, consider what a particular reader or group of readers needs and whether generated images are satisfactory, or discovering links to pre-existing, human-created images and video is necessary. We leave a discussion of automatic manipulation of academic figures and tables to future work.

\subsection{Good Quality Footnotes}
In addition to the variety of types of footnotes, participants also frequently discussed what qualities make a \textit{good} or \textit{useful} footnote, including credible, balanced sources, reasonable footnote length, filtered results, and unobtrusive and passive presentation.

\subsubsection{Sources}
It is important to be aware of which sources a system is allowed to consider. Quality as it relates to sources mostly focuses on the credibility of the source.

The system needs a way to determine how to prioritize the information it has access to.

In short, the sources that the AI uses need to be credible and balanced. What counts as credible may vary by use case. It needs to simultaneously avoid false and useless information but not exclude opinions simply in an attempt to tell the user what they want to hear. This may be the most important consideration when choosing what AI algorithms to use and how to compose the database of sources, if such a database is necessary. Much scholarship has gone into looking at the biases of ChatGPT~\cite{chatgpt-bias}. And if an LLM is chosen as the main footnote generation algorithm, a thorough understanding of the biases and vagaries of that particular LLM is crucial to producing quality footnotes. 

\subsubsection{Conception}
We see a variation of this idea implemented in Google search results in the section ``People also ask,'' which allows users to click on a related question, evaluate the snippet of information provided, and decide if they want to follow a link to further information. A similar idea could potentially be implemented here to obviate the need for the algorithm to be extremely precise in its ability to present exactly the right amount of information, relying instead on some means for the reader to indicate their desire for further information.

The main things to be considered in formulating footnotes, then, are making sure that the information is both relevant and accurate, and that enough information is given to the reader to inform them without overwhelming them. Some participants mentioned wanting a complete list of sources in some contexts, while wanting a curated list in others, particularly in instances where a complete list may be overwhelming.

The information the system provides needs to be simultaneously relevant and accurate and adhere to the Goldilocks principle. The length of the footnote needs to be not too short, not too long, and how much information that actually is may vary on a case-by-case basis.

This might be at least partially mitigated by allowing the reader to control the flow of information, initially giving them smaller amounts of information and allowing them to request further information as necessary, like a footnote within a footnote.



\subsubsection{Presentation}
Allowing a user free access to use AI footnotes when they want them and allowing them to turn off the feature when not useful might go a long way to avoiding the fate of some user interface agents such as Microsoft Word's Clippy~\cite{clippy}.


\subsection{No One-Size-Fits-All}
A common thread throughout many of the themes presented is the idea that one solution may not be possible for each situation. Instead, it may be better for designers to choose a context and design for readers/writers in that specific context, rather than trying to provide a catchall solution for every situation. Users reading for more academic purposes would likely need sources sampled from a different population of documents than a user reading a novel for pleasure.

\section{Limitations}

Our participants tend to be very educated. While this is useful in the sense that they tend to have a lot of experience with using footnotes, they are not representative of a more general population.
In building an actual AI footnote system, it would be important to consider how well the needs of the users of a particular AI footnote system align with the needs outlined here, and if there are other aspects of the implementation that need to be considered. 

In addition, AI research continues apace. ChatGPT and OpenAI have continued to release new features (ChatGPT was often referenced by our participants). If we conducted these interviews again, it is likely we might see many newly available AI features reflected in suggestions for AI footnotes.


\section{Conclusion \& Future Work}
We have provided a conceptual foundation to help those who wish to build a functional AI footnote system. Future work will certainly involve implementing a functional AI footnote system and deploying it with readers. 

There also appeared to be variability between how participants wanted to have alternative perspectives presented to them in different contexts. Particularly in religious contexts, several participants mentioned not wanting to be shown information that was outside of their own religious perspective--specifically perspectives that directly contradicted their religious beliefs. We did not explore this avenue further, but it may be that individuals want an AI system that will not interfere with their personal moral code. Further research could investigate this and other areas of life where users are uncomfortable with AI intruding.

By interviewing current footnote users and analyzing their responses, we provide a lens through which practitioners can evaluate reader desires when implementing an AI footnote system. Using this lens may help them make informed choices based on the needs of their readers and the resources available to their footnote system.



\begin{acks}

\end{acks}

\bibliographystyle{ACM-Reference-Format}
\bibliography{sample-base}

\appendix

\end{document}